\documentclass[aps,prb,preprint,groupedaddress,showpacs]{revtex4}

\usepackage{graphicx}
\usepackage{amsmath}

\newcommand{\RDC}{%
  \ensuremath{R_{\mathrm{DC}}}}
\newcommand{\Rph}{%
  \ensuremath{R_{\mathrm{ph}}}}
\newcommand{\Rlw}{%
  \ensuremath{R_{\mathrm{985}}}}
\newcommand{\Ict}{%
  \ensuremath{I_{c,t}}}
\newcommand{\Ics}{%
  \ensuremath{I_{c,s}}}

\begin{document}

\title{Magnetic-field dependence of count rates in superconducting thin-film TaN single-photon detectors}

\author{A.\ Engel} \email[]{andreas.engel@physik.uzh.ch}
\author{A.\ Schilling}
\affiliation{Physics Institute of the University of Zurich, Winterthurerstr. 190, 5087 Zurich, Switzerland}
\author{K.\ Il'in}
\author{M.\ Siegel}
\affiliation{Institute of Micro- and Nano-Electronic Systems, Karlsruhe Institute of Technology, Hertzstr.\ 16, 76187 Karlsruhe, Germany}

\date{\today}

\begin{abstract}
We have studied the magnetic-field dependence of both dark-count rates and photon-count rates in a superconducting nanowire single-photon detector made of TaN in external magnetic fields $\left|\mu_0H\right| < 10$~mT perpendicular to the plane of the underlying meander structure and at $T = 4$~K. The dark-count rates show a characteristic field-dependence, which is asymmetric with respect to magnetic field direction. The field- and the current dependence of the dark counts can be quantitatively well explained if one assumes that the critical current is reduced to $\approx50\%$ at the $180^\circ$ meander turns when compared to a straight strip, and the observed asymmetry can be modeled assuming that the turnarounds are not all strictly equal. Surprisingly, the photon count rates do not show any significant field dependence, which seems to be at odds with existing detection models invoking vortex crossings.
\end{abstract}


\maketitle



Magnetic vortices in meso- and microscopic superconducting structures display a rich variety of phenomena. Since in many applications such structures are composed of arrays of superconducting strips, the strip geometry is of particular interest. It has been realized theoretically \cite{Clem98a,Maksimova98} and experimentally \cite{Stan04} that narrow superconducting strips remain in the Meissner state up to magnetic fields significantly exceeding the lower critical field $H_{c1}$ of the bulk material. However, if a transport current is applied, vortices (or anti-vortices) may be activated over the energy barrier at the strip edges that subsequently traverse the strip under the influence of the Lorentz-force, thereby destroying the dissipation-free superconducting state \cite{Tafuri06}. Currently there is a debate\cite{Qiu08,Bulaevskii11,Vodolazov12,Bulaevskii12,Gurevich12} over the fluctuation modes dominating in thin superconducting strips that are too wide to be considered one-dimensional, \emph{i.e.}, with width $w\gg\xi$, the superconducting coherence length, but narrow enough so that edge-effects are significant, \emph{i.e.}\ $w\ll\Lambda$, where $\Lambda=2\lambda^2/d$ is the effective penetration depth, $\lambda$ is the magnetic penetration depth in thick films and $d\ll \lambda$ is the film thickness. This situation is typical for superconducting nanowire single-photon detectors\cite{Goltsman01} (SNSPD), where a large bias current is applied which is of the order of the experimental critical current. Vortices crossing the superconducting strip have been considered as the main cause for dark-count events\cite{Bartolf10, Bulaevskii11} in SNSPD. Vortices have also been made responsible to explain the detection of low-energy photons for which the photon energy alone is insufficient to trigger a detection event \cite{Hofherr10,Bulaevskii12}. Both events, dark counts as well as vortex-assisted photon counts, should have a distinct magnetic-field dependence \cite{Bulaevskii12}.

The detection elements of SNSPD are not simply straight strips, however. In order to increase the effective detection area and the coupling to the electro-magnetic wave of the photons, the active area typically consists of a meander with separations between adjacent strips roughly equal to the strip width. This design implies $180^\circ$ turnarounds at the strip ends for the superconducting bias-current. Whereas the current-density along the straight portions of the meander is highly uniform, due to the fact that $w\ll\Lambda$, the current density is inhomogeneous in the vicinity of the turnarounds\cite{Clem11}. The local current density at the inner radius of the turnaround exceeds the uniform current density of the straight parts and it is reduced at the outer radius. This leads to an effectively reduced critical current for a meander structure, because superconductivity is destroyed when the energy barrier vanishes at the inner radius. For the geometry considered here, the corresponding current density is always lower than the depairing current-density\cite{Bulaevskii11}. The level of reduction of the critical current depends on the detailed geometry of the turn \cite{Hortensius12,Henrich12} and can be minimized to achieve higher critical currents for meander structures\cite{Akhlaghi12a}. In the case of a single turn or several turns in the same direction, the application of a small magnetic field of the right orientation should lead to an increase of the critical current\cite{Clem12}.


In this Rapid Communication we report on the first systematic measurements of the magnetic field-dependence of dark counts as well as photon counts in SNSPD. The device studied in detail is made from a $d=4.9$~nm thick TaN film deposited by DC reactive magnetron sputtering on a R-plane cut sapphire substrate\cite{Ilin12}. Using electron-beam lithography the film was patterned into a meander with $21$ strips of width $w=110$~nm, strip separation $\Delta=80$~nm and covering an area of $3.4\times3.9$~$\mu$m$^2$. Such TaN-SNSPD have been demonstrated to be single-photon detectors comparing favorably well with NbN-SNSPD\cite{Engel12}. After fabrication the device had a sharp superconducting transition at $T_c=8.84$~K.
Count-rate measurements were done in a He-3 cryostat with free-space optical access. The light from a continuous light-source was passed through a monochromator and directed onto the detector through a series of quartz windows and apertures with decreasing diameter to reduce the amount of blackbody background radiation. For the measurements of the intrinsic dark counts caused by fluctuations in the superconducting meander, the optical windows were replaced by blanks and an Al cap was placed over the detector in good thermal contact to the sample holder, thereby creating an isothermal cavity around the detector.

The bias current was supplied by a \emph{Keithley 2400} programmable source-measure unit. To ensure a stable DC current through the detector, a series of HF- and low-pass filters was used. The detector signal was fed into a cryogenic amplifier at the 4K-stage and then further amplified at room-temperature before fed into a threshold counter. Further details of the used set-up are given elsewhere\cite{Engel12}. A bipolar magnetic field up to about $10$~mT was applied using a conventional electro-magnet placed as close as possible to the device outside the cryostat and opposite to the optical access window. In this way the magnetic field was oriented perpendicular to the meander plane. The magnetic field was measured using a calibrated cryogenic hall-sensor that was placed in close proximity to the detector inside the cryostat. We verified field gradients to be negligible over the small detector area due to a minimum distance of the magnet to the detector of about $6$~cm\footnote{Using a dipolar approximation for the field, the relative variations over the detector area are $<10^{-7}$ and the in-plane component is estimated $<1\%$, even for moderate misalignments.}. All measurements presented have been done at $T=4$~K.

\begin{figure}
 \includegraphics[width=\columnwidth,totalheight=200mm,keepaspectratio]{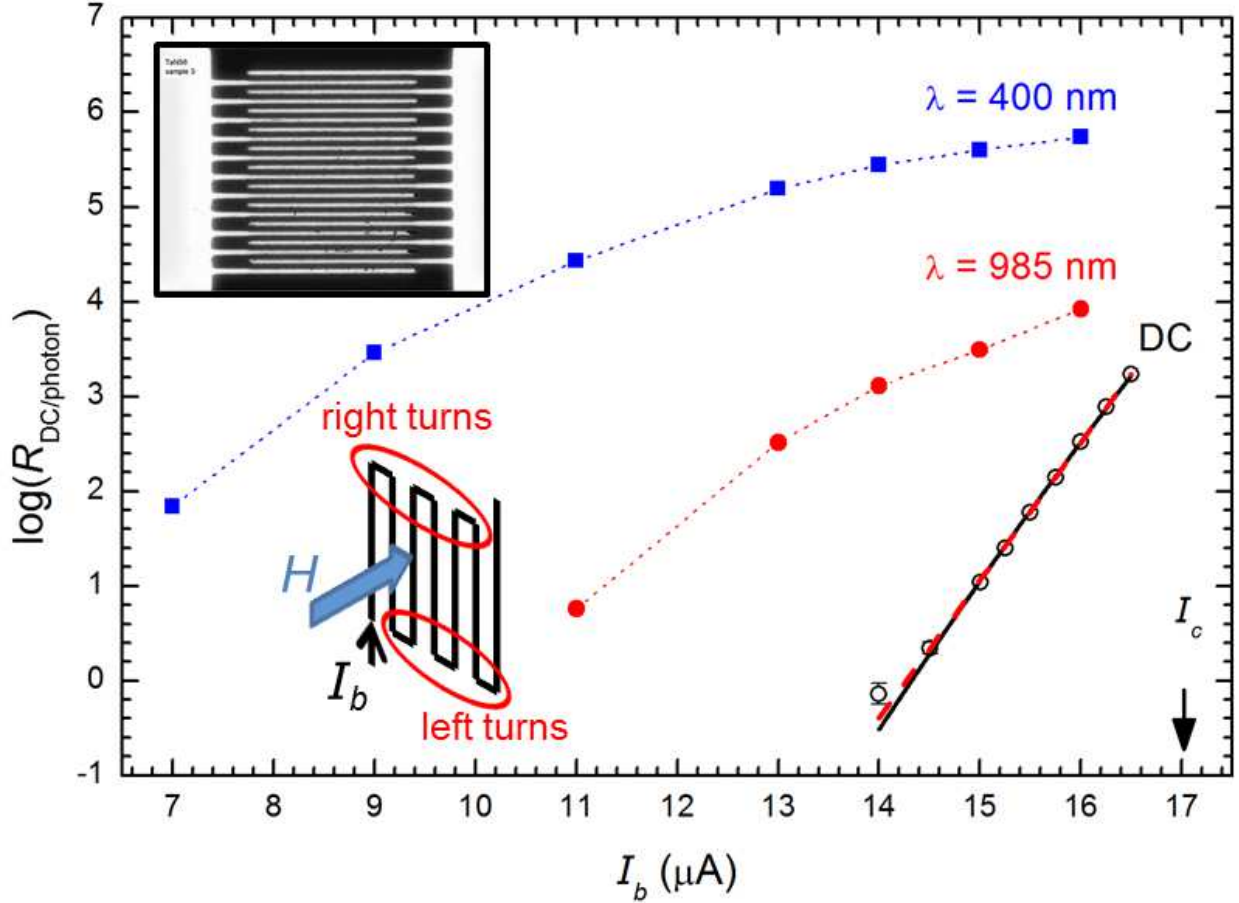}%
 \caption{Dark counts and photon counts on logarithmic scale \emph{vs}. bias current. Errors are of the size of the data points or smaller except where plotted. The solid line is a least-squares fit to Eq.~(51) from Ref.~\onlinecite{Bulaevskii11}, while the dashed line is a fit to the phenomenological Eq.~\eqref{Eq.RDC}. The dashed lines connecting the photon-counts data are guides to the eye. The inset shows an electron micrograph of the measured device. In the lower left corner is a sketch of the magnetic field and bias current in relation to the meander geometry.\label{Fig.DCbias}}
\end{figure}

In Fig.\ \ref{Fig.DCbias} the dark-count rate \RDC\ in zero magnetic field is plotted on a logarithmic scale \emph{vs.}\ the applied bias current $I_b$. It shows the typical near-exponential increase for increasing bias currents. The solid line is a fit of Eq.~(51) from Ref.~\onlinecite{Bulaevskii11} to our data with an attempt rate and the vortex energy scale $\nu$ as free parameters (solid line in Fig.\ \ref{Fig.DCbias}). We obtain an almost identical current-dependence upon reversing the current direction, but shifted along the $I_b$-axis by about $0.4~\mu$A, which is probably caused by an unaccounted current offset (not shown in Fig.~\ref{Fig.DCbias}). From the fitted parameter $\nu=\varepsilon_0/(k_BT)=51.5\pm1.5$, $\varepsilon_0=\Phi_0^2\mu^2/(2\pi\mu_0\Lambda)$ (Refs.\ \onlinecite{Bulaevskii11,Bulaevskii12}), we can derive the effective penetration depth at $T=4$~K, $\Lambda=165\pm5$~$\mu$m, which compares very well with $\sim120~\mu$m, the zero-$T$ value expected from our conductivity measurements, where $\Phi_0=2.068\times10^{-15}$~Wb is the magnetic-flux quantum, $\mu=0.930$ a correction for currents close to the depairing current, $\mu_0=4\pi\times10^{-7}$ Vs/Am is the vacuum permeability and $k_B=1.38\times10^{-23}$~J/K is Boltzmann's constant.

For the description of the magnetic-field dependence of the dark-count rate we can phenomenologically approximate the current-dependence as
\begin{equation}
\RDC \approx R_0\exp\left(\frac{I_b}{\tilde{I}}\right),\label{Eq.RDC}
\end{equation}
with $R_0$ being a proportionality factor, $\tilde{I}\propto I_c$ is a current scale and $I_c$ is the experimental critical current. The dashed line in Fig.\ \ref{Fig.DCbias} as a resulting fit of our data to Eq.\ \eqref{Eq.RDC} shows that for the current-range of interest this is a satisfactory description of the current-dependence of \RDC. Using Eq.\ \eqref{Eq.RDC} has the advantage over Eq.~(51) of Ref.~\onlinecite{Bulaevskii11} that the underlying physical model, which we will use to describe the field-dependence, becomes more transparent. In the following we will assume that the experimental critical current $I_c$ is equivalent to the current for which the barrier for vortex entry vanishes at a certain point along an edge of our structure. The experimentally measured $I_c\approx17~\mu$A is indicated in Fig.~\ref{Fig.DCbias}.

\begin{figure}
 \includegraphics[width=\columnwidth,totalheight=200mm,keepaspectratio]{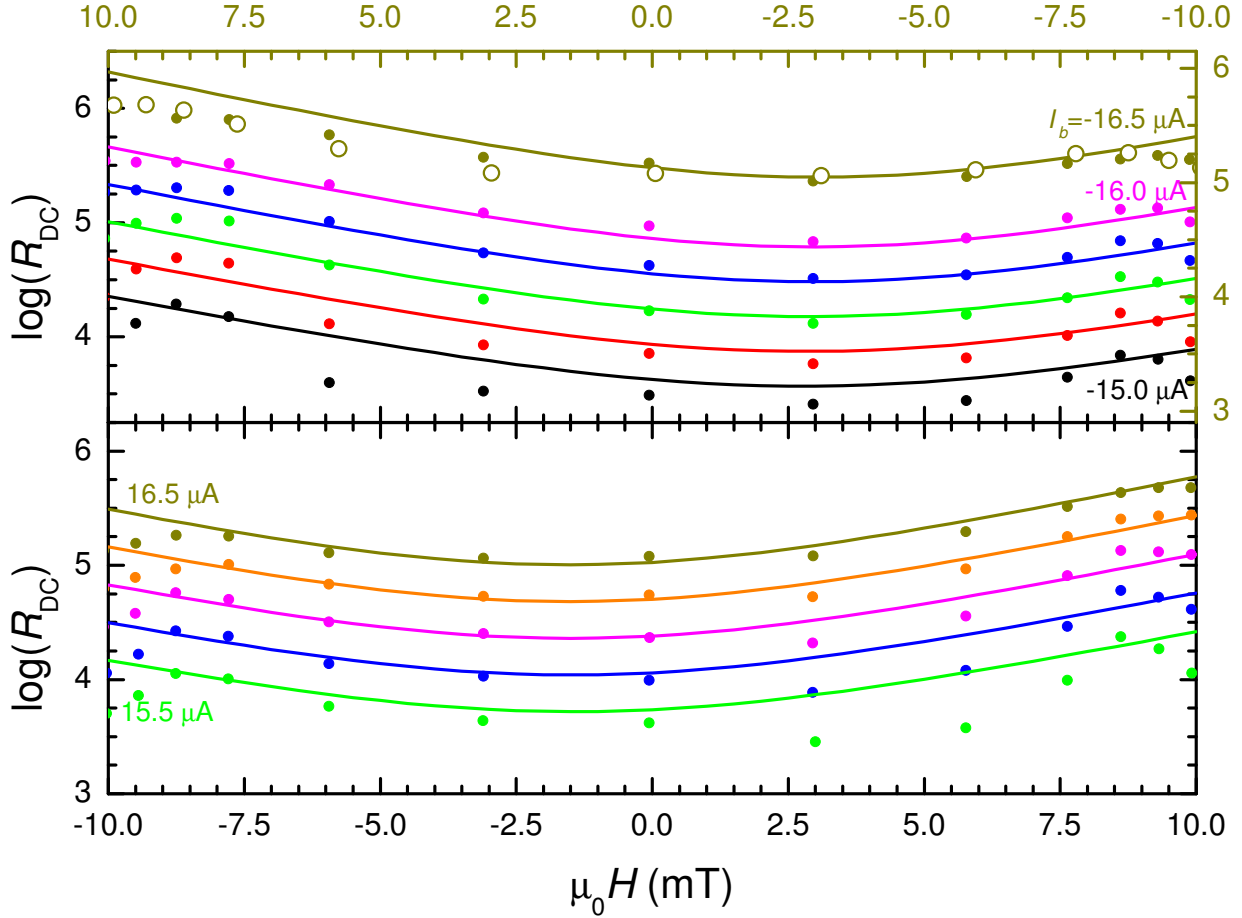}%
 \caption{Dark counts $\log(\RDC)$ for different $I_b$ plotted as a function of applied magnetic field $H$. The $\left|I_b\right|$ have been progressively increased by $0.25~\mu$A. Upper panel: $I_b<0$; lower panel: $I_b>0$. The $+16.5~\mu$A data are also plotted in the upper panel, but on a reversed field-axis (uppermost $H$-axis) to demonstrate the equivalence of reversing field or current direction. The corresponding y-axis (right axis, upper panel) has been shifted by $0.35$ to compensate the current off-set. Solid lines are least-square fits according to Eq.~\eqref{Eq.Meander} with common parameters, performed independently for negative and positive $I_b$, respectively.\label{Fig.DCfield}}
\end{figure}

In Fig.\ \ref{Fig.DCfield} we are plotting the measured \RDC\ as a function of the applied magnetic field $H$ for various bias currents. The upper panel shows the dark-count rates for a certain current direction (in our convention $I_b<0$) and the lower panel for the reversed current direction ($I_b>0$), both for positive and negative magnetic-field values. The dark-count rate is clearly magnetic-field dependent with a certain unexpected asymmetry with respect to the field direction. For $I_b>0$ ($I_b<0$) the minimum in \RDC\ seems to be shifted to slightly negative (positive) fields. However, reversing both the current and magnetic field direction leaves the asymmetry unchanged as demonstrated for the data measured with $I_b=+16.5~\mu$A plotted in the upper panel (open symbols), but with a reversed (upper) field-axis from $+10$ to $-10$~mT.

Following a recent theoretical model for vortex-crossings in straight strips\cite{Bulaevskii12} we would expect a symmetric field-dependence of the dark-count rate around $H=0$, $\RDC(H)\propto\cosh(H/H_1)$, with $H_1$ being a magnetic-field scale for vortex crossings. This behavior is clearly not compatible with the measured data, even if taking into account that the meander geometry consists of several parallel straight strips with alternating current directions. The self-field of one strip amounts to at most about $150~\mu$T at the edges and the contributions from all the neighboring strips partly cancel each other, resulting in a maximum contribution of $\approx\pm20~\mu$T, which is too small to explain the observed asymmetry in $\RDC(H)$.

A more plausible explanation is that the $180^\circ$ turnarounds connecting neighboring strips are responsible for the asymmetry in the magnetic-field dependence, which we will demonstrate in the following.

In general, a single $180^\circ$ turnaround is expected to have a reduced critical current \Ict\ as compared to the straight strip (\Ics)\cite{Clem12}. Applying the equations for the sharp rectangular $180^\circ$ turnaround to our geometry, we calculate a reduction $\alpha=\Ict/\Ics\approx0.5$. Estimating the critical current of the straight sections $\Ics=2wI_0/(\pi e\xi)$, with $I_0=\Phi_0\mu^2/(2\mu_0\Lambda)$ at $4$~K, and using the T-dependence of $\Lambda$ and $\xi$ and $\xi(0)\approx5$~nm from $H_{c2}(T)$ measurements\cite{Bartolf10}, we have $\alpha\approx17\ \mu\text{A}/22.6\ \mu\text{A}\approx0.75$. The measured $\alpha$ is larger than the theoretical value, most probably because our turnarounds are not strictly rectangular, but somewhat rounded.

Furthermore, it becomes clear that for bias currents $I_b\lesssim \Ict$, the barriers for vortex-entry along the straight sections of the meander are still relatively high. Using Eq.\ \eqref{Eq.RDC} and scaling $\tilde{I}$ with $I_c$ we estimate the probability for vortex-crossings near the turnaround to be a factor $\sim10^5$ higher than along the straight sections, making vortex-crossings near the turnarounds the dominating contribution to \RDC\ despite their short overall length.

The application of a magnetic field destroys the symmetry of the current distribution in left- and right-turning turnarounds. In one case (e.g.\ left turn) screening currents due to the applied field will increase the current density at the inner edge (lower experimental $I_c$), while in the other case (right turn) it will decrease the current density (higher $I_c$). For magnetic fields $H\ll H^\ast$, with $H^\ast$ being a device-dependent field scale for the critical current, the critical current is linear in magnetic field\cite{Bulaevskii12,Clem12},
\begin{equation}
I_c(H)\approx\Ict\left(1-\frac{H}{H^\ast}\right),\ H^\ast=\frac{2w}{\pi e\xi}H_0,\ H_0=\frac{\Phi_0}{2\mu_0w^2},\label{Eq.IcH}
\end{equation}
with \Ict\ the critical current in zero-field, and we obtain $\mu_0 H^\ast\approx425$~mT for our device.

In Ref.~\onlinecite{Bulaevskii12}, the magnetic-field dependence of the dark-count rate was obtained by replacing $I_b$ with the true microscopic current at the edge which is the sum of applied current and screening current. We will take here a complementary approach by identifying the critical current for which the barrier for vortex entry vanishes with the field-dependent critical current from Eq.~\eqref{Eq.IcH}\cite{Clem12}. Inserting Eq.~\eqref{Eq.IcH} into \eqref{Eq.RDC} results in the field-dependent dark-count rate for a single $180^\circ$ turnaround
\begin{equation}
\RDC^1(H)=R_0\exp\left(\frac{I_b}{\tilde{I}(1-H/H^\ast)}\right).\label{Eq.SingleTurn}
\end{equation}
The application of a positive magnetic field would then lead to a reduction of the critical current and thus an increased dark-count rate, whereas a negative field leads to a reduction of $\RDC^1(H)$. Considering the geometry of our meander (see Fig.~\ref{Fig.DCbias}), we have an equal number of left- and right-turns with opposite effects of positive and negative applied fields on $I_c$, and the overall effect of a magnetic-field should again be symmetric around $H=0$. In reality, however, we may not expect that all of the turns are identical due to imperfections from the structuring process, and we have to assume a certain distribution of $\alpha$-factors. In order to keep the model as simple as possible, we assume one turn to have a reduced critical-current (by a factor $\beta\leq1$) and all others to have equal \Ict. We then arrive at the following equation for the field-dependent dark-count rate, neglecting the  contributions from the straight sections with $\Ics>\Ict$,
\begin{equation}
\RDC^M(H)=(N-1)\RDC^1(H)+\RDC^\beta(H)+N\RDC^1(-H),\label{Eq.Meander}
\end{equation}
with $\RDC^\beta(H)=R_0\exp\left[I_b/\left(\beta\tilde{I}(1-H/H^\ast)\right)\right]$ and $N=10$ the number of left and right turns, respectively for the measured device. We performed independent least-squares fits to our data obtained for positive and negative currents (solid lines in Fig.~\ref{Fig.DCfield}), with $\mu_0H^\ast\approx240$~mT, $\beta=0.94\pm0.01$ and $\tilde{I}=0.35\pm0.01~\mu$A  as additional fitting parameters to the attempt rate $R_0$. $\tilde{I}$ compares well with $\tilde{I}=0.31\pm0.01~\mu$A from fits of Eq.~\eqref{Eq.RDC} to the zero-field $\RDC(I_b)$ data. The field-scale $H^\ast$ is smaller than the expected value calculated with Eq.\ \eqref{Eq.IcH} possibly due to edge imperfections generally leading to reduced edge barriers. However, our simple, vortex-based model obviously fairly accurately describes both the magnitude of and the observed asymmetry in the measured $\RDC(H)$ with reasonable physical parameters.


\begin{figure}
 \includegraphics[width=\columnwidth,totalheight=200mm,keepaspectratio]{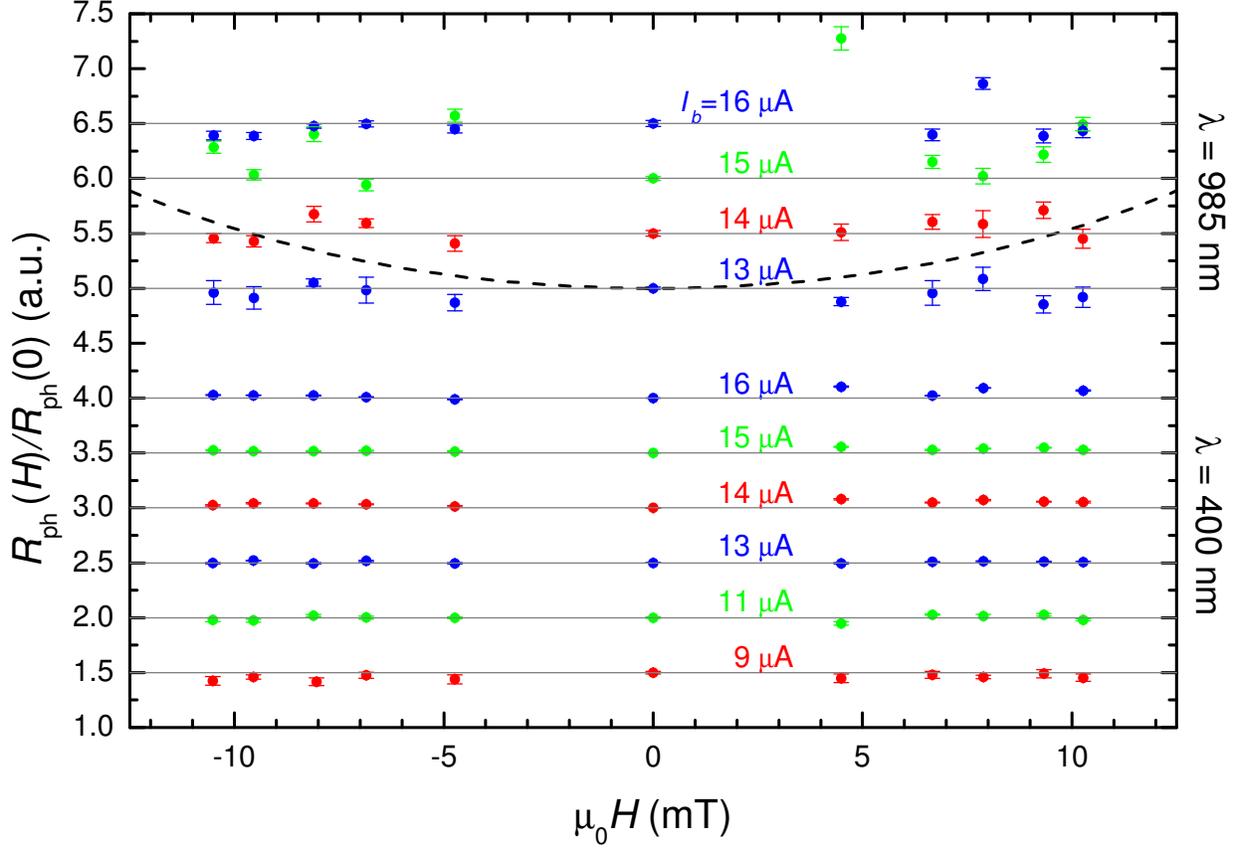}%
 \caption{Relative photon count-rates as functions of the applied magnetic field $H$, normalized to photon count-rates in zero-field for each $I_b$. For clarity, the  datasets have been shifted in $y$-direction by multiples of $0.5$. We also plotted the expected $H$-dependence for $985$~nm photons and $I_b=13\ \mu$A (dashed black curve). Horizontal lines are to guide the eye.\label{Fig.PhotonsField}}
\end{figure}

In Fig.~\ref{Fig.PhotonsField} we present our measurements of photon count-rates \Rph\ as functions of the applied magnetic field, normalized to the photon count-rates in zero-field and for two photon wavelengths of $400$~nm and $985$~nm, respectively. Dark-counts (typically $\ll0.1\Rph$, compare Fig.\ \ref{Fig.DCbias}) have been subtracted. The estimated detection efficiencies (DE) range from $\sim0.1$ at maximum $I_b$ and $400$~nm photons, to $\approx10^{-4}$ at the lowest measured $I_b$ and $985$~nm photons. From the DE up to $0.1$ we can conclude that most photon-detection events occur along the straight sections of the meander\cite{Ilin12}, which is also supported by recent simulations \cite{Berdiyorov12a}. Except for the scattering in some datasets, \Rph\ appear to be constant with a \emph{field-independent} photon count-rate.

Next, we will estimate the expected field-dependence following Ref.\ \onlinecite{Bulaevskii12} for the photon count-rate at $985$~nm wavelength, \Rlw, and $I_b=13\ \mu$A, for which the field-dependence should be most pronounced. Assuming that for $400$~nm photons and $I_b=16\ \mu$A we have reached the maximum $\mathrm{DE}\approx R_\mathrm{abs}$, the photon-absorption rate, we obtain for the photon-assisted vortex-crossing rate $\frac{\Rlw(13\ \mu\mathrm{A})}{R_\mathrm{abs}}=1-\exp(-2\eta)=6\cdot10^{-4}\approx\eta$ (Eq.\ (45) in [\onlinecite{Bulaevskii12}])\footnote{A factor of 2 is missing in Eq.\ (45) in [\onlinecite{Bulaevskii12}], since vortices and anti-vortices contribute to the count-rate.}. With $\eta\ll1$ Eq.\ (47) in [\onlinecite{Bulaevskii12}] can be simplified to
\begin{equation}
\frac{\Rlw(H)}{\Rlw(0)}\approx\cosh\left(\frac{H}{H_{1h}}\right),\label{Eq.PhotonField}
\end{equation}
with $H_{1h}=H^\ast\left(\nu_h+1\right)^{-1}I_b/\Ics$, and $\nu_h$ is the vortex energy-scale near the absorption site that can be estimated from the low-current part in Fig.\ \ref{Fig.DCbias}, where $\Rph\propto I^{\nu_h}$, see Eq.\ (46) in [\onlinecite{Bulaevskii12}]. We estimate $\nu_h\approx12$ and therefore, using the larger, theoretical $H^\ast$, we obtain $\mu_0H_{1h}\approx10$~mT at $I_b=13\ \mu$A. The expected field-dependence is plotted in Fig.\ \ref{Fig.PhotonsField} (dashed curve) and would be easily measurable, which is clearly inconsistent with our data, however. A similarly strong field dependence would be expected for $400$~nm photons and $I_b =9\ \mu$A ($\mu_0H_{1h}\approx12.5$~mT), whereas $H_{1h}$ for the other datasets is expected to be larger and the field-dependence less pronounced, yet should still be measurable.

We note here that we have obtained qualitatively similar
results also in NbN-SNSPD (not shown here), which shows that our observations are not unique to the TaN material


In conclusion, we have presented systematic measurements of dark-count rates and photon-count rates as a function magnetic field in a TaN-SNSPD. The field-dependence of $\RDC(H)$ can be successfully modeled assuming vortices traversing the superconducting strip predominantly in the vicinity of the meander turns, where the critical current for vortex-entry is reduced with respect to the straight sections. The observed asymmetry in \RDC\ with respect to current- or field-reversal can be explained assuming that not all turnarounds have exactly the same \Ict. By contrast, we have not found any evidence for a scenario in which vortices assist in the detection of photons with energies that are insufficient to trigger a detection event. If vortices contribute, the current model obviously significantly underestimates the relevant field-scales, or certain peculiarities of the meander geometry may have to be taken into account. Our results are not only relevant for the application of SNSPD, but should be of great interest for a broad range of situations in which magnetic vortices in small-scale structures are subject to external fields and currents.

We acknowledge stimulating discussions with L.~Bulaevskii and A.~Semenov. This research received support from the Swiss National Science Foundation grant No.\ 200021\_135504/1 and is supported in part by DFG Center for Functional Nanostructures under sub-project A4.3.

\bibliography{Literature}

\begin{thebibliography}{20}
\expandafter\ifx\csname natexlab\endcsname\relax\def\natexlab#1{#1}\fi
\expandafter\ifx\csname bibnamefont\endcsname\relax
  \def\bibnamefont#1{#1}\fi
\expandafter\ifx\csname bibfnamefont\endcsname\relax
  \def\bibfnamefont#1{#1}\fi
\expandafter\ifx\csname citenamefont\endcsname\relax
  \def\citenamefont#1{#1}\fi
\expandafter\ifx\csname url\endcsname\relax
  \def\url#1{\texttt{#1}}\fi
\expandafter\ifx\csname urlprefix\endcsname\relax\def\urlprefix{URL }\fi
\providecommand{\bibinfo}[2]{#2}
\providecommand{\eprint}[2][]{\url{#2}}

\bibitem[{\citenamefont{Clem}(1998)}]{Clem98a}
\bibinfo{author}{\bibfnamefont{J.~R.} \bibnamefont{Clem}}, in
  \emph{\bibinfo{booktitle}{Bull. Am. Phys. Soc.}} (\bibinfo{year}{1998}),
  vol.~\bibinfo{volume}{43}, p. \bibinfo{pages}{411}.

\bibitem[{\citenamefont{Maksimova}(1998)}]{Maksimova98}
\bibinfo{author}{\bibfnamefont{G.~M.} \bibnamefont{Maksimova}},
  \bibinfo{journal}{Phys. Solid State} \textbf{\bibinfo{volume}{40}},
  \bibinfo{pages}{1607} (\bibinfo{year}{1998}).

\bibitem[{\citenamefont{Stan et~al.}(2004)\citenamefont{Stan, Field, and
  Martinis}}]{Stan04}
\bibinfo{author}{\bibfnamefont{G.}~\bibnamefont{Stan}},
  \bibinfo{author}{\bibfnamefont{S.~B.} \bibnamefont{Field}}, \bibnamefont{and}
  \bibinfo{author}{\bibfnamefont{J.~M.} \bibnamefont{Martinis}},
  \bibinfo{journal}{Phys. Rev. Lett.} \textbf{\bibinfo{volume}{92}},
  \bibinfo{pages}{097003} (\bibinfo{year}{2004}).

\bibitem[{\citenamefont{Tafuri et~al.}(2006)\citenamefont{Tafuri, Kirtley,
  Born, Stornaiuolo, Medaglia, Orgiani, Balestrino, and Kogan}}]{Tafuri06}
\bibinfo{author}{\bibfnamefont{F.}~\bibnamefont{Tafuri}},
  \bibinfo{author}{\bibfnamefont{J.~R.} \bibnamefont{Kirtley}},
  \bibinfo{author}{\bibfnamefont{D.}~\bibnamefont{Born}},
  \bibinfo{author}{\bibfnamefont{D.}~\bibnamefont{Stornaiuolo}},
  \bibinfo{author}{\bibfnamefont{P.~G.} \bibnamefont{Medaglia}},
  \bibinfo{author}{\bibfnamefont{P.}~\bibnamefont{Orgiani}},
  \bibinfo{author}{\bibfnamefont{G.}~\bibnamefont{Balestrino}},
  \bibnamefont{and} \bibinfo{author}{\bibfnamefont{V.~G.} \bibnamefont{Kogan}},
  \bibinfo{journal}{Europhys.\ Lett.} \textbf{\bibinfo{volume}{73}},
  \bibinfo{pages}{948} (\bibinfo{year}{2006}).

\bibitem[{\citenamefont{Qiu and Qian}(2008)}]{Qiu08}
\bibinfo{author}{\bibfnamefont{C.}~\bibnamefont{Qiu}} \bibnamefont{and}
  \bibinfo{author}{\bibfnamefont{T.}~\bibnamefont{Qian}},
  \bibinfo{journal}{Phys. Rev. B} \textbf{\bibinfo{volume}{77}},
  \bibinfo{pages}{174517} (\bibinfo{year}{2008}).

\bibitem[{\citenamefont{Bulaevskii et~al.}(2011)\citenamefont{Bulaevskii, Graf,
  Batista, and Kogan}}]{Bulaevskii11}
\bibinfo{author}{\bibfnamefont{L.~N.} \bibnamefont{Bulaevskii}},
  \bibinfo{author}{\bibfnamefont{M.~J.} \bibnamefont{Graf}},
  \bibinfo{author}{\bibfnamefont{C.~D.} \bibnamefont{Batista}},
  \bibnamefont{and} \bibinfo{author}{\bibfnamefont{V.~G.} \bibnamefont{Kogan}},
  \bibinfo{journal}{Phys. Rev. B} \textbf{\bibinfo{volume}{83}},
  \bibinfo{pages}{144526} (\bibinfo{year}{2011}), \eprint{1102.5130}.

\bibitem[{\citenamefont{Vodolazov}(2012)}]{Vodolazov12}
\bibinfo{author}{\bibfnamefont{D.~Y.} \bibnamefont{Vodolazov}},
  \bibinfo{journal}{Phys. Rev. B} \textbf{\bibinfo{volume}{85}},
  \bibinfo{pages}{174507} (\bibinfo{year}{2012}), \eprint{1204.0365}.

\bibitem[{\citenamefont{Bulaevskii et~al.}(2012)\citenamefont{Bulaevskii, Graf,
  and Kogan}}]{Bulaevskii12}
\bibinfo{author}{\bibfnamefont{L.~N.} \bibnamefont{Bulaevskii}},
  \bibinfo{author}{\bibfnamefont{M.~J.} \bibnamefont{Graf}}, \bibnamefont{and}
  \bibinfo{author}{\bibfnamefont{V.~G.} \bibnamefont{Kogan}},
  \bibinfo{journal}{Phys. Rev. B} \textbf{\bibinfo{volume}{85}},
  \bibinfo{pages}{014505} (\bibinfo{year}{2012}), \eprint{1108.4004}.

\bibitem[{\citenamefont{Gurevich and Vinokur}(2012)}]{Gurevich12}
\bibinfo{author}{\bibfnamefont{A.}~\bibnamefont{Gurevich}} \bibnamefont{and}
  \bibinfo{author}{\bibfnamefont{V.~M.} \bibnamefont{Vinokur}},
  \bibinfo{journal}{Phys. Rev. B} \textbf{\bibinfo{volume}{86}},
  \bibinfo{pages}{026501} (\bibinfo{year}{2012}), \eprint{1201.5347}.

\bibitem[{\citenamefont{Gol'tsman et~al.}(2001)\citenamefont{Gol'tsman, Okunev,
  Chulkova, Lipatov, Semenov, Smirnov, Voronov, Dzardanov, Williams, and
  Sobolewski}}]{Goltsman01}
\bibinfo{author}{\bibfnamefont{G.~N.} \bibnamefont{Gol'tsman}},
  \bibinfo{author}{\bibfnamefont{O.}~\bibnamefont{Okunev}},
  \bibinfo{author}{\bibfnamefont{G.}~\bibnamefont{Chulkova}},
  \bibinfo{author}{\bibfnamefont{A.}~\bibnamefont{Lipatov}},
  \bibinfo{author}{\bibfnamefont{A.}~\bibnamefont{Semenov}},
  \bibinfo{author}{\bibfnamefont{K.}~\bibnamefont{Smirnov}},
  \bibinfo{author}{\bibfnamefont{B.}~\bibnamefont{Voronov}},
  \bibinfo{author}{\bibfnamefont{A.}~\bibnamefont{Dzardanov}},
  \bibinfo{author}{\bibfnamefont{C.}~\bibnamefont{Williams}}, \bibnamefont{and}
  \bibinfo{author}{\bibfnamefont{R.}~\bibnamefont{Sobolewski}},
  \bibinfo{journal}{Appl. Phys. Lett.} \textbf{\bibinfo{volume}{79}},
  \bibinfo{pages}{705} (\bibinfo{year}{2001}).

\bibitem[{\citenamefont{Bartolf et~al.}(2010)\citenamefont{Bartolf, Engel,
  Schilling, Il'in, Siegel, H\"ubers, and Semenov}}]{Bartolf10}
\bibinfo{author}{\bibfnamefont{H.}~\bibnamefont{Bartolf}},
  \bibinfo{author}{\bibfnamefont{A.}~\bibnamefont{Engel}},
  \bibinfo{author}{\bibfnamefont{A.}~\bibnamefont{Schilling}},
  \bibinfo{author}{\bibfnamefont{K.}~\bibnamefont{Il'in}},
  \bibinfo{author}{\bibfnamefont{M.}~\bibnamefont{Siegel}},
  \bibinfo{author}{\bibfnamefont{H.-W.} \bibnamefont{H\"ubers}},
  \bibnamefont{and} \bibinfo{author}{\bibfnamefont{A.}~\bibnamefont{Semenov}},
  \bibinfo{journal}{Phys. Rev. B} \textbf{\bibinfo{volume}{81}},
  \bibinfo{pages}{024502} (\bibinfo{year}{2010}).

\bibitem[{\citenamefont{Hofherr et~al.}(2010)\citenamefont{Hofherr, D., Ilin,
  Siegel, Semenov, H\"{u}bers, and Gippius}}]{Hofherr10}
\bibinfo{author}{\bibfnamefont{M.}~\bibnamefont{Hofherr}},
  \bibinfo{author}{\bibfnamefont{R.}~\bibnamefont{D.}},
  \bibinfo{author}{\bibfnamefont{K.}~\bibnamefont{Ilin}},
  \bibinfo{author}{\bibfnamefont{M.}~\bibnamefont{Siegel}},
  \bibinfo{author}{\bibfnamefont{A.}~\bibnamefont{Semenov}},
  \bibinfo{author}{\bibfnamefont{H.-W.} \bibnamefont{H\"{u}bers}},
  \bibnamefont{and} \bibinfo{author}{\bibfnamefont{N.~A.}
  \bibnamefont{Gippius}}, \bibinfo{journal}{J. Appl. Phys.}
  \textbf{\bibinfo{volume}{108}}, \bibinfo{eid}{014507}
  (pages~\bibinfo{numpages}{9}) (\bibinfo{year}{2010}).

\bibitem[{\citenamefont{Clem and Berggren}(2011)}]{Clem11}
\bibinfo{author}{\bibfnamefont{J.~R.} \bibnamefont{Clem}} \bibnamefont{and}
  \bibinfo{author}{\bibfnamefont{K.~K.} \bibnamefont{Berggren}},
  \bibinfo{journal}{Phys. Rev. B} \textbf{\bibinfo{volume}{84}},
  \bibinfo{pages}{174510} (\bibinfo{year}{2011}), \eprint{1109.4881}.

\bibitem[{\citenamefont{Hortensius et~al.}(2012)\citenamefont{Hortensius,
  Driessen, Klapwijk, Berggren, and Clem}}]{Hortensius12}
\bibinfo{author}{\bibfnamefont{H.~L.} \bibnamefont{Hortensius}},
  \bibinfo{author}{\bibfnamefont{E.~F.~C.} \bibnamefont{Driessen}},
  \bibinfo{author}{\bibfnamefont{T.~M.} \bibnamefont{Klapwijk}},
  \bibinfo{author}{\bibfnamefont{K.~K.} \bibnamefont{Berggren}},
  \bibnamefont{and} \bibinfo{author}{\bibfnamefont{J.~R.} \bibnamefont{Clem}},
  \bibinfo{journal}{Appl. Phys. Lett.} \textbf{\bibinfo{volume}{100}},
  \bibinfo{pages}{182602} (\bibinfo{year}{2012}), \eprint{1203.4253}.

\bibitem[{\citenamefont{Henrich et~al.}(2012)\citenamefont{Henrich,
  Reichensperger, Hofherr, Il'in, Siegel, Semenov, Zotova, and
  Vodolazov}}]{Henrich12}
\bibinfo{author}{\bibfnamefont{D.}~\bibnamefont{Henrich}},
  \bibinfo{author}{\bibfnamefont{P.}~\bibnamefont{Reichensperger}},
  \bibinfo{author}{\bibfnamefont{M.}~\bibnamefont{Hofherr}},
  \bibinfo{author}{\bibfnamefont{K.}~\bibnamefont{Il'in}},
  \bibinfo{author}{\bibfnamefont{M.}~\bibnamefont{Siegel}},
  \bibinfo{author}{\bibfnamefont{A.}~\bibnamefont{Semenov}},
  \bibinfo{author}{\bibfnamefont{A.}~\bibnamefont{Zotova}}, \bibnamefont{and}
  \bibinfo{author}{\bibfnamefont{D.~Y.} \bibnamefont{Vodolazov}}
  (\bibinfo{year}{2012}), \eprint{1204.0616}.

\bibitem[{\citenamefont{Akhlaghi et~al.}(2012)\citenamefont{Akhlaghi, Atikian,
  Eftekharian, Loncar, and Majedi}}]{Akhlaghi12a}
\bibinfo{author}{\bibfnamefont{M.~K.} \bibnamefont{Akhlaghi}},
  \bibinfo{author}{\bibfnamefont{H.}~\bibnamefont{Atikian}},
  \bibinfo{author}{\bibfnamefont{A.}~\bibnamefont{Eftekharian}},
  \bibinfo{author}{\bibfnamefont{M.}~\bibnamefont{Loncar}}, \bibnamefont{and}
  \bibinfo{author}{\bibfnamefont{A.~H.} \bibnamefont{Majedi}}
  (\bibinfo{year}{2012}), \eprint{1205.4290}.

\bibitem[{\citenamefont{Clem et~al.}(2012)\citenamefont{Clem, Mawatari,
  Berdiyorov, and Peeters}}]{Clem12}
\bibinfo{author}{\bibfnamefont{J.~R.} \bibnamefont{Clem}},
  \bibinfo{author}{\bibfnamefont{Y.}~\bibnamefont{Mawatari}},
  \bibinfo{author}{\bibfnamefont{G.~R.} \bibnamefont{Berdiyorov}},
  \bibnamefont{and} \bibinfo{author}{\bibfnamefont{F.~M.}
  \bibnamefont{Peeters}}, \bibinfo{journal}{Phys. Rev. B}
  \textbf{\bibinfo{volume}{85}}, \bibinfo{pages}{144511}
  (\bibinfo{year}{2012}), \eprint{1111.5233}.

\bibitem[{\citenamefont{Il'in et~al.}(2012)\citenamefont{Il'in, Hofherr, Rall,
  Siegel, Semenov, Engel, Inderbitzin, Aeschbacher, and Schilling}}]{Ilin12}
\bibinfo{author}{\bibfnamefont{K.}~\bibnamefont{Il'in}},
  \bibinfo{author}{\bibfnamefont{M.}~\bibnamefont{Hofherr}},
  \bibinfo{author}{\bibfnamefont{D.}~\bibnamefont{Rall}},
  \bibinfo{author}{\bibfnamefont{M.}~\bibnamefont{Siegel}},
  \bibinfo{author}{\bibfnamefont{A.}~\bibnamefont{Semenov}},
  \bibinfo{author}{\bibfnamefont{A.}~\bibnamefont{Engel}},
  \bibinfo{author}{\bibfnamefont{K.}~\bibnamefont{Inderbitzin}},
  \bibinfo{author}{\bibfnamefont{A.}~\bibnamefont{Aeschbacher}},
  \bibnamefont{and}
  \bibinfo{author}{\bibfnamefont{A.}~\bibnamefont{Schilling}},
  \bibinfo{journal}{J. Low Temp. Phys.} \textbf{\bibinfo{volume}{167}},
  \bibinfo{pages}{809} (\bibinfo{year}{2012}).

\bibitem[{\citenamefont{Engel et~al.}(2012)\citenamefont{Engel, Aeschbacher,
  Inderbitzin, Schilling, Il'in, Hofherr, Siegel, Semenov, and
  H\"{u}bers}}]{Engel12}
\bibinfo{author}{\bibfnamefont{A.}~\bibnamefont{Engel}},
  \bibinfo{author}{\bibfnamefont{A.}~\bibnamefont{Aeschbacher}},
  \bibinfo{author}{\bibfnamefont{K.}~\bibnamefont{Inderbitzin}},
  \bibinfo{author}{\bibfnamefont{A.}~\bibnamefont{Schilling}},
  \bibinfo{author}{\bibfnamefont{K.}~\bibnamefont{Il'in}},
  \bibinfo{author}{\bibfnamefont{M.}~\bibnamefont{Hofherr}},
  \bibinfo{author}{\bibfnamefont{M.}~\bibnamefont{Siegel}},
  \bibinfo{author}{\bibfnamefont{A.}~\bibnamefont{Semenov}}, \bibnamefont{and}
  \bibinfo{author}{\bibfnamefont{H.-W.} \bibnamefont{H\"{u}bers}},
  \bibinfo{journal}{Appl. Phys. Lett.} \textbf{\bibinfo{volume}{100}},
  \bibinfo{pages}{062601} (\bibinfo{year}{2012}).

\bibitem[{\citenamefont{Berdiyorov et~al.}(2012)\citenamefont{Berdiyorov,
  Milo\v{s}evi\'{c}, and Peeters}}]{Berdiyorov12a}
\bibinfo{author}{\bibfnamefont{G.~R.} \bibnamefont{Berdiyorov}},
  \bibinfo{author}{\bibfnamefont{M.~V.} \bibnamefont{Milo\v{s}evi\'{c}}},
  \bibnamefont{and} \bibinfo{author}{\bibfnamefont{F.~M.}
  \bibnamefont{Peeters}} (\bibinfo{year}{2012}), \eprint{1206.4298}.

\end{thebibliography}

\end{document}